\documentclass{article}

\usepackage{PRIMEarxiv}

\usepackage[utf8]{inputenc} 
\usepackage[T1]{fontenc}    
\usepackage{hyperref}       
\usepackage{url}            
\usepackage{booktabs}       
\usepackage{amsfonts}       
\usepackage{nicefrac}       
\usepackage{microtype}      
\usepackage{lipsum}
\usepackage{fancyhdr}       
\usepackage{graphicx}       
\graphicspath{{media/}}     

\usepackage{algorithm}
\usepackage{algorithmic}
\usepackage{bbm}
\usepackage{amsmath}
\usepackage{amsthm}

\usepackage{placeins}

\usepackage{caption}
\usepackage{subcaption}
\usepackage{xr}
\newtheorem{lemma}{Lemma}
\newtheorem{theorem}{Theorem}
\newtheorem{corollary}{Corollary}

\usepackage{tikz}
\usetikzlibrary{bayesnet}
\usepackage{ amsfonts, amssymb}
\tikzset{>=latex}

\newcommand{\calW}{\mathcal{W}}
\newcommand{\calC}{\mathcal{C}}
\newcommand{\calY}{\mathcal{Y}}
\newcommand{\calX}{\mathcal{X}}
\title{A Light-weight, Effective and Efficient Model for Label Aggregation in Crowdsourcing
}

\author{
  Yi Yang, Zhong-Qiu Zhao\\
  Hefei University of Technology \\
  Hefei, China\\
  \texttt{\{yyang, z.zhao\}@hfut.edu.cn} \\
   \And
  Quan Bai \\
  University of Tasmania \\
  Hobart, Australia\\
  \texttt{Quan.Bai@utas.edu.au} \\
   \AND
   Qing Liu \\
   Data61, CSIRO  \\
   Hobart, Australia \\
   \texttt{q.liu@data61.csiro.au} \\
   \And
   Weihua Li \\
   Auckland University of Technology \\
   Auckland, New Zealand \\
   \texttt{weihua.li@aut.ac.nz} \\
}

\begin{document}
\maketitle

\begin{abstract}
Due to the noises in crowdsourced labels, label aggregation (LA) has emerged as a standard procedure to post-process crowdsourced labels. LA methods estimate true labels from crowdsourced labels by modeling worker qualities. Most existing LA methods are iterative in nature. They need to traverse all the crowdsourced labels multiple times in order to iteratively update true labels and worker qualities until convergence. Consequently, these methods have high space complexity $\mathcal{O}(TM)$ and time complexity $\mathcal{O}(ITM)$, where $T$ and $M$ are the numbers of tasks and workers, respectively, and $I$ is the number of iterations these algorithms take to converge. In this paper, we treat LA as a dynamic system and model it as a Dynamic Bayesian network. From the dynamic model we derive two light-weight algorithms, LA\textsuperscript{onepass} and LA\textsuperscript{twopass}, which can effectively and efficiently estimate worker qualities and true labels by traversing all the labels at most twice. As a result,  the space and time complexities of the proposed algorithms are reduced to $\mathcal{O}(M+T)$ and $\mathcal{O}(MT)$, respectively. Due to the dynamic nature and low complexities, the proposed algorithms can also estimate true labels online without re-visiting historical data. We theoretically prove the convergence of the proposed algorithms, and bound the error of estimated worker qualities. Experiments conducted on 20 real-world datasets demonstrate that the proposed algorithms can effectively and efficiently aggregate labels both offline and online even if they traverse all the labels at most twice.
\end{abstract}


\section{Introduction}
Many tasks of machine learning require labels for training. Traditional label collection from domain experts or data vendors is usually expensive and time-consuming, which may not meet the increasing demand of labels. As an alternative, crowdsourcing is economical and efficient for acquiring labels \cite{howe2008crowdsourcing,callison2010creating}. Crowdsourcing platforms, such as Amazon Mechanic Turk \cite{ipeirotis2010analyzing} and FigureEight \cite{de2015using}, help label requesters to distribute labeling tasks to the crowd workers who will label the distributed tasks and receive some monetary reward. Despite the low cost of crowdsourcing, the crowd workers are not experts and may erroneously label some tasks \cite{kang2021crowdsourcing}. Consequently, the labels acquired by crowdsourcing are usually less accurate than those from experts. Therefore, it is common to collect multiple labels for each task from different workers, and aggregate the collected labels to alleviate errors \cite{jin2020technical}.

Aggregating crowdsourced labels is referred to as label aggregation (LA) or truth inference in crowdsourcing \cite{zheng2017truth} or truth discovery in the database community \cite{li2016survey}. LA takes the crowdsourced labels as input and estimates the true label for each task. Typically, LA is unsupervised because there are no ground-truth labels available for supervision. The most straightforward LA algorithm is majority voting (MV) \cite{franklin2011crowddb}. MV simply regards the label received the most votes from workers as the true label for each task. Due to its simplicity, MV's space and time complexities are as low as $\mathcal{O}(M+T)$ and $\mathcal{O}(TM)$ where $M$ is the number of crowd workers and $T$ is the number of tasks.  However, MV assumes all the workers are equally reliable, which is typically false in crowdsourcing \cite{jin2020technical}. To address this limitation, recent LA methods choose to model worker qualities while estimating true labels. The underlying assumption is that workers who often label tasks correctly are of high quality, while the label supported by high quality workers is selected as the true label for each task. Guided by this assumption, existing LA algorithms estimate true labels and worker qualities jointly and iteratively until some convergence condition is satisfied \cite{dawid1979maximum,kim2012bayesian,li2019exploiting,yang2021unsupervised}. Various empirical results demonstrate that LA methods modeling worker quality generally outperform MV in terms of accuracy \cite{zheng2017truth}. However, most existing LA methods have two limitations due to their iterative nature. First, they need to load the entire dataset (crowdsourced labels) into memory, which has space complexity at least as high as $\mathcal{O}(TM)$. Second, they need to iteratively traverse the entire dataset multiple times, whose time complexity is at least as high as $\mathcal{O}(ITM)$ where $I$ denotes the number of iterations the algorithms take to converge.


In this paper, we develop a light-weight, effective and efficient LA algorithm, named LA\textsuperscript{onepass}, that has lower space and time complexities than that of iterative algorithms. Specifically, we tag each label a time-slice, which is the index of the label's task. As a result, the labels and tasks have temporal attributes. We then treat LA as a dynamic system, where worker qualities evolve after estimating true labels over time. We use Dynamic Bayesian network \cite{koller2009probabilistic} to model such dynamic system, where worker qualities are modeled as (unknown) temporal variables that evolve over time, while the (observed) labels and (unknown) true labels are modeled as non-temporal variables that are only instantiated within one time-slice. When estimating the unknown variables at every time-slice, the worker qualities can be estimated efficiently by Maximum A Posterior (MAP), and the true label can be estimated by solving a simple optimization problem analytically. This ensures the crowdsourced labels are traversed only once. It reduces the space and time complexities down to $\mathcal{O}(M + T)$ and $\mathcal{O}(TM)$, which equal to the space and time complexities of MV. We also prove that the estimated worker quality converges, and the rate of convergence is also given. Moreover, the error of estimated worker quality can be bounded with high probability even if the crowdsourced labels are traversed only once.

Traversing the labels only once inherently has one disadvantage. The true labels that are estimated early may not be accurate because the estimated worker qualities have not converged. Therefore, we develop LA\textsuperscript{twopass}, an extension of LA\textsuperscript{onepass}, that uses the estimated worker qualities from  LA\textsuperscript{onepass} to re-estimate the true labels by performing weighted majority voting. LA\textsuperscript{twopass} traverses the crowdsourced labels twice, but can substaitianlly improve aggragation accuracy. LA\textsuperscript{twopass} has little computational overhead compared to LA\textsuperscript{onepass} because it does not estimate worker qualities when it traverses the labels again. Due to low space and time complexities, both LA\textsuperscript{onepass} and LA\textsuperscript{twopass} can also be configured to aggregate labels online.

We perform experiments on 20 real-world datasets to demonstrate the effectiveness and efficiency of the proposed algorithms compared to state-of-the-art LA methods, and show the estimated worker qualities converge and its error can be bounded by simulated experiments.

\section{Problem Statement \& Related Work}
In this section, we first formally define the problem of LA. Then we review the prior work of LA from three dimensions that are related to our work.

\subsection{Problem Statement of Label Aggregation}
Suppose there are $T$ tasks and $M$ workers. Each task $t$ has $K$ (mutual exclusive) classes indexed from $1$ to $K$, and its unknown true label $y_t$ is drawn from $[K]$ where $[K]$ denotes the set of integers $\{1,\dots,K\}$. The tasks are labeled by the workers, and the label from worker $i \in [M]$ for task $t \in [T]$ is denoted as  $x_{i,t}$, where $x_{i,t} \in [K]$. Each worker $i$ is associated with an unknown variable $w_i$ reflecting the quality of the worker's labels. For convenience, we denote the whole crowdsourced labels as $\calX = \{{x_{i,t} | i \in [M], t \in [T]}\}$.  The goal is to aggregate $\calX$ and estimate the true labels $\calY = \{\hat{y}_t | t \in [T]\}$ for every task as well as the worker qualities $\calW = \{\hat{w}_i | i \in [M]\}$. For notation simplicity, we assume each worker labels all the tasks. However, the proposed algorithms can also deal with label sparsity when each worker labels a subset of tasks.

Please note that we study single-truth LA problem in this paper, where the true label of a task has only one value. This is different from multi-truth LA problems \cite{zhang2019multi}. Our method is also universal, which only takes crowdsourced labels as input. This also differs from LA methods that additionally take features of workers or tasks as input. For example, \cite{yang2021unsupervised} uses workers' social network in their aggregation model.  \cite{liu2021aggregating} uses side information to cluster tasks. However, these features are difficult to obtain in crowdsourcing and are not available in every crowdsourcing applications.

\subsection{Prior Work}
\textbf{Worker quality modeling. } Active research works of LA focus on modeling worker quality. The first LA method considering worker quality can be dated back to 1979 when Dawid and Skene proposed an algorithm, commonly known as DS, to aggregate clinical diagnoses of doctors \cite{dawid1979maximum}. DS is classified as the ``confusion matrix'' model in the literature, because it uses a $K \times K$ confusion matrix to capture the probabilities that a worker's label is generated from a task conditioned on the task's true label. A large number of LA methods are descendants of DS \cite{kim2012bayesian,zhang2014spectral,hong2021online}. For example, LFC \cite{raykar2010learning} extends DS by adding priors to confusion matrices. EBCC \cite{li2019exploiting} learns worker correlations when estimating the confusion matrices.  Another commonly adopted worker quality model is ``one-coin'' model. One-coin model treats the quality of each worker as a single parameter reflecting the quality of worker's labels. For example, ZC \cite{demartini2012zencrowd} models the worker quality as a value between $0$ and $1$ representing the probability of a worker's label being correct. IWMV \cite{li2014error} transforms such probability for estimating true labels which has provable theoretical guarantee on the error rate.  There are also some one-coin models treating worker quality as a real number, where higher value means the worker's labels are more likely to be true \cite{li2014resolving,li2019truth}. 

\textbf{Modeling techniques and solution framework.} Probabilistic graphical model (PGM) \cite{koller2009probabilistic} is the most widely used technique to solve LA problem \cite{whitehill2009whose,kim2012bayesian,li2019truth,jin2020technical}. PGM depicts conditional dependencies between random variables. Most PGMs used in LA are generative, which model the conditional probability of a worker's label given the unknown worker quality and true label. Techniques other than PGM are also used. The optimization-based method \cite{li2014resolving} directly builds an objective function capturing the relation between worker qualities and true labels. \cite{ma2020gradient} constructs the crowd labels as a matrix and estimates worker qualities and true labels by matrix completion. Methods using neural network, such as LAA \cite{li2017aggregating}, are also used in LA by modeling the non-linear relationship among crowd labels, worker qualities and true labels. 

Regardless of the modeling techniques, most derived LA algorithms are iterative, which can be summarized in Algorithm \ref{alg1} \cite{zheng2017truth}. It can be observed that Algorithm \ref{alg1} iteratively estimates true labels and worker qualities by traversing the entire dataset $\calX$ multiple times as described in the \textit{while} loop. 

\begin{algorithm}[tb]
	\caption{LA general solution framework}
	\label{alg1}
	\textbf{Input}: $\calX$\\
	\textbf{Output}: $\calY$ and  $\calW$
	\begin{algorithmic}[1] 
		\STATE Initialize worker qualities $\calW$
		\WHILE{Not Converged} 
		\FOR{$t \in [T]$}
		\STATE Estimate true label $\hat{y}_t$ based on $\calX$ and $\calW$;
		\ENDFOR
		\FOR{$i \in [M]$}
		\STATE Estimate worker quality $\hat{w}_i$ based on $\calX$ and $\calY$;
		\ENDFOR
		\ENDWHILE
		\STATE \textbf{return} $\calY = \{\hat{y}_t | t \in [T]\}$, $\calW = \{\hat{w}_i | i \in [M]\}$
	\end{algorithmic}
\end{algorithm}

\textbf{Online LA. } Online LA refers to the ability to aggregate labels at present without re-visiting historical labels  when labels are arrived in chunks sequentially \cite{hong2021online}. Online LA was studied in the database community when labels are of continuous and are passively collected from data sources such as webs \cite{li2015discovery, yang2019probabilistic,yang2019dynamic}. However, in the crowdsourcing applications, the labels are typically categorical. To the best of authors' knowledge, the methods that can aggregate labels online in crowdsourcing are SBIC \cite{manino2019streaming} and BiLA \cite{hong2021online}. However, SBIC is limited in applicability because it can only aggregate labels of tasks with two classes. BiLA uses neural network to represent its internal probability distributions. It requires to traverse the present labels multiple times in order to update worker qualities, which is inefficient as demonstrated in the experiments. In contrast, the proposed algorithms in this paper can aggregate labels of tasks with any number of classes, and traverse the labels at most twice to accurately and efficiently aggregate labels online.

\section{Method}
In this section, we describe the proposed Dynamic Bayesian network (DBN) model for LA, and present LA\textsuperscript{onepass} algorithm derived from the proposed dynamic model.

\subsection{The Dynamic Model}
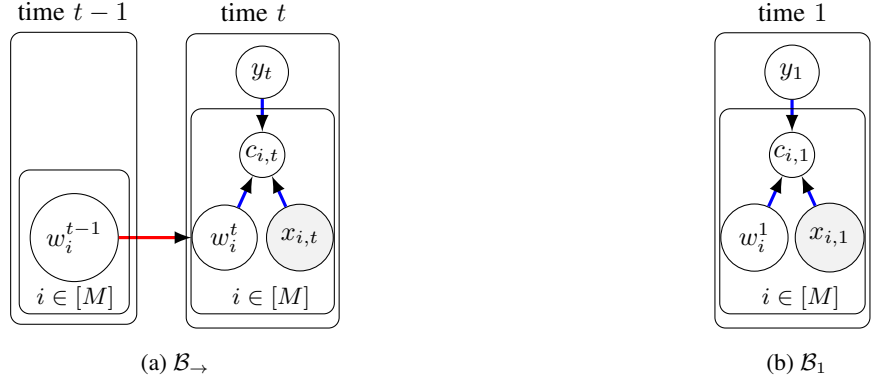
\begin{figure}[!ht]
	\centering
	\begin{subfigure}[b]{0.66\columnwidth}
		\centering
		\begin{tikzpicture}
			\node[circle,draw=black,fill=white!10,inner sep=0pt,minimum size=0.6cm] (cnext) at (2,-1) {$c_{i,t}$};
			\node[circle,draw=black,fill=gray!10,,minimum size=0.4cm] (xnext) at (2.5,-2.1) {$x_{i,t}$};
			\node[circle,draw=black,fill=white!10,,minimum size=0.6cm] (wnext) at (1.5,-2.1) {$w_i^{t}$};
			\node[circle,draw=black,fill=white!10,,minimum size=0.6cm] (ynext) at (2,0.1) {$y_{t}$};
			
			
			\node[circle,draw=black,fill=white!10,,minimum size=1.0cm] (wnow) at (-0.5,-2.1) {$w_{i}^{t-1}$};
			\node[mark size=1pt,color=black] (emp) at (0.3,0.4) {$ $};
			
			\node [text width=1.2cm] (m) at (2.2,-2.9) {\small{$i \in [M]$}};
			\node [text width=1.2cm] (m2) at (-0.4,-2.9) {\small{$i \in [M]$}};
			
			\path[draw=blue,very thick,->] (wnext) edge (cnext);
			\path[draw=blue,very thick,->] (xnext) edge (cnext);
			\path[draw=blue,very thick,->] (ynext) edge (cnext);
			
			\path[draw=red,very thick,->] (wnow) edge (wnext);
			
			\node [text width=1.5cm] (t1) at (2.2,0.9) {time $t$};
			\node [text width=1.5cm] (t1) at (-0.5,0.9) {time $t-1$};
			
			\plate[inner sep=.0cm,yshift=0.3cm] {plate1} {(wnext)(xnext) (cnext)  (m) } { };
			\plate[inner sep=0.05cm,yshift=0.1cm] {plate2} {(cnext)(xnext) (wnext) (ynext) (plate1.west) (plate1.east) (plate1.south)} { };
			\plate[inner sep=.0cm,xshift=-0.1cm,yshift=.3cm] {plate3} {(wnow)(m2)} { };
			
			\plate[inner sep=.0cm,xshift=-0.1cm,yshift=0.1cm] {plate4} {(wnow)(emp) (plate3.west) (plate3.east) (plate3.south)} { };

		\end{tikzpicture}
		\caption{$\mathcal{B_\rightarrow}$}
	\end{subfigure}
	\begin{subfigure}[b]{0.33\columnwidth}
		\centering
		\begin{tikzpicture}
			\node[circle,draw=black,fill=white!20,inner sep=0pt,minimum size=0.6cm] (cnext) at (2,-1) {$c_{i,1}$};
			\node[circle,draw=black,fill=gray!10,,minimum size=0.6cm] (xnext) at (2.5,-2.1) {$x_{i,1}$};
			\node[circle,draw=black,fill=white!10,,minimum size=0.6cm] (wnext) at (1.5,-2.1) {$w_i^{1}$};
			\node[circle,draw=black,fill=white!10,,minimum size=0.6cm] (ynext) at (2,0.1) {$y_{1}$};
			

			\node [text width=1.2cm] (m) at (2.2,-2.9) {\small{$i \in [M]$}};
			
			\path[draw=blue,very thick,->] (wnext) edge (cnext);
			\path[draw=blue,very thick,->] (xnext) edge (cnext);
			\path[draw=blue,very thick,->] (ynext) edge (cnext);

			\node [text width=1.5cm] (t1) at (2.3,0.9) {time $1$};
			
			\plate[inner sep=.0cm,yshift=0.3cm] {plate1} {(wnext)(xnext) (cnext) (m) } { };
			\plate[inner sep=0.05cm,yshift=0.1cm] {plate2} {(cnext)(xnext) (wnext) (ynext) (plate1.west) (plate1.east) (plate1.south)} { };

		\end{tikzpicture}
		\caption{$\mathcal{B}_1$}
	\end{subfigure}

	\caption{The proposed DBN model. The intra-time-slice edge is in red; the inter-time-slice edge is in blue. }
	\label{fig_dbn}
\end{figure}

We treat LA as a dynamic system evolving over $T$ discrete time-slices, which equal to the number of tasks. In the dynamic system, the meaning of $t$ is two-fold: it represents the index of task $t$ and the time-slice, or state, of the system is at. Likewise, $T$ represents the number of tasks as well as the life span of the system. At each time-slice $t \in [T]$, it only estimates task $t$'s true label. We model worker quality $w_i \in [0,1]$ reflecting the probability that this worker's label is true. The worker quality $w_i$ evolves over time, and is updated each time after $y_t$ is estimated until it reaches the end at time-slice $T$. 

The dynamic system can be modeled by a DBN. In the DBN, $w_i$ is modeled as a temporal variable evolving over time. We add a superscript to the worker quality $w_i^t$ to denote its state at time-slice $t$. $x_{i,t}$ and $y_t$ are modeled as non-temporal variables that are only instantiated within the their own time-slice. The DBN can be described by two Bayesian networks $\mathcal{B}_\rightarrow$ and $\mathcal{B}_1$ as shown in Figure \ref{fig_dbn}. $\mathcal{B}_\rightarrow$ is 2-time-slice Bayesian network (2TBN) depicting the relation of variables within one time-slice and the evolution of variables in between two consecutive time-slices. $\mathcal{B}_1$ is a Bayesian network depicting the initial state of the system.

\subsubsection{2TBN $\mathcal{B}_\rightarrow$. } There are two types of edges, \textit{inter-time-slice} edge and  \textit{intra-time-slice} edge, connecting variables in $\mathcal{B}_\rightarrow$. The inter-time-slice edges connect variables $w_i^t$, $x_{i,t}$ and $y_t$ within time-slice $t$. The relation between these variables is expressed via an auxiliary deterministic variable $c_{i,t}$ indicating whether worker $i$ labels task $t$ correctly. Therefore, $c_{i,t} = \mathbbm{1}(x_{i,t} = y_t)$, where $\mathbbm{1}(\cdot)$ is the indicator function. Since the correctness of worker $i$'s label is determined by $w_i^t \in [0,1]$, we can model $c_{i,t}$ as a Bernoulli random variable with parameter $w_i^t$:
\begin{align}
	c_{i,t} \sim Ber(w_i^t),\hspace{0.2cm } p(c_{i,t}|w_i^t) = (w_i^t)^{c_{i,t}} (1-w_i^t)^{1-c_{i,t}}
\end{align}

The intra-time-slice edge connects $w_i^{t-1}$ and $w_i^t$. This edge expresses how a worker quality evolves over time by specifying the transition probability $p(w_i^t|w_i^{t-1})$. We treat $p(w_i^t|w_i^{t-1})$ as the posterior distribution of $w_i^{t-1}$ after observing the labels of task $t-1$ and its true label has been estimated. In other words, $w_i^{t-1}$ is the prior of $w_i^t$.

\subsubsection{Initial state $\mathcal{B}_1$. } $\mathcal{B}_1$ expresses the initial state of the dynamic system. The relation of variables is the same as that of time-slice $t$ in $\mathcal{B}_\rightarrow$, except that it needs to specify the initial state of temporal variable $w_i^1$. Since $w_i$ is the probability that worker $i$ labels tasks correctly, we model $w_i$ as a Beta random variable and its initial state $w_i^1$ is associated with a Beta distribution having hyperparameters $\alpha$ and $\beta$:
\begin{align}
	w_i^1 \sim Beta(\alpha,\beta), \hspace{0.2cm} p(w_i^1) \propto (w_i^1)^{\alpha - 1} (1-w_i^1)^{\beta - 1}
\end{align}

\subsection{Estimation}
Given the model architecture, the joint probability of $\calW_t = \{w_i^t | i \in [M]\}$, $\calC_t=\{c_{i,t}|i \in [M]\}$ and $\calX_t = \{x_{i,t}| i \in [M]\}$ within  time-slice $t$ can be factorized as 
\begin{align}
	\begin{split}
		p(\calW_t, \calC_t,\calX_t,y_t) &= \prod_{i} p(c_{i,t}|w_i^t,x_{i,t},y_t)\\
		& = \prod_i (w_i^t)^{c_{i,t}} (1-w_i^t) ^{1-c_{i,t}} 
	\end{split}
\end{align}
and its log-likelihood function $l_t = \log p(\calW_t, \calC_t,\calX_t,y_t)$ is 
\begin{align}
	\begin{split}
		&l_t  = \sum_{i=1}^M c_{i,t} \log w_i^t + (1-c_{i,t}) \log(1-w_i^t) \\
		& = \sum_{i=1}^M \mathbbm{1}(x_{i,t} = y_t) \log w_i^t + (1-\mathbbm{1}(x_{i,t} = y_t)) \log(1-w_i^t)
	\end{split}
\end{align}

We estimate the true label $y_t$ by maximizing $l_t$, which can be easily solved via: 
\begin{align} \label{eq_yt}
	\hat{y}_t = \arg\max_k \{ \sum_{i=1}^M w_i^t \mathbbm{1}(x_{i,t} = k) | k \in [K] \}
\end{align}

Provided that $w_i^{t-1}$ is the prior of $w_i^t$, the posterior probability of $w_i^t$ after observing $x_{i,t}$ and estimating $y_t$ is
\begin{align*}
	\begin{split}
		&p(w_i^{t} | c_{i,t}, x_{i,t},y_t,w_i^{t-1})\propto p(c_{i,t}| w_i^t, x_{i,t},y_t) p(w_i^t|w_i^{t-1}) 
	\end{split}
\end{align*}
By the chain rule of probability, the above posterior probability can be expanded as:
\begin{align}
	\begin{split}
		&p(w_i^{t} | c_{i,t},w_i^{t-1}) \propto \prod_{t'=1}^t p(c_{i,t'}| w_i^{t'}) p(w_i^1) \\
		& = (w_i^{t})^{ C_{i,t} + \alpha - 1} (1-w_i^{t})^{t - C_{i,t}+ \beta - 1}
	\end{split}
\end{align}
where $C_{i,t} = \sum_{t'=1}^t\mathbbm{1}(x_{i,t'} = \hat{y}_{t'})$ is the number of tasks worker $i$ has labeled correctly up to time-slice $t$, and $t - C_{i,t}$ is the number of tasks worker $i$ has labeled incorrectly up to time-slice $t$. It can be observed  that the posterior $p(w_i^{t} | c_{i,t},w_i^{t-1})$ can be compactly written as $p(w_i^t | C_{i,t})$, and  $ p(w_i^t | C_{i,t}) \sim Beta (C_{i,t} + \alpha, t - C_{i,t}+ \beta)$, which is again a Beta distribution. Therefore, we can estimate $w_i^t$ by Maximum a Posteriori (MAP):
\begin{align} \label{eq_wit}
	\hat{w}_i^t = \frac{C_{i,t} + \alpha - 1}{t + \alpha + \beta -2}                             
\end{align}
The form of Equation (\ref{eq_wit}) is expected. It reflects the estimated probability that worker $i$ labels tasks correctly up to time-slice $t$ with the prior belief $\alpha$ and $\beta$.

\subsubsection{Algorithm flow.} Given the estimation in Equations (\ref{eq_yt}) and (\ref{eq_wit}), the LA algorithm derived from the proposed DBN is summarized in Algorithm \ref{alg}. Since all the crowdsourced labels are passed to the algorithm only once, we term the algorithm LA\textsuperscript{onepass}.

\begin{algorithm}[tb]
	\caption{LA\textsuperscript{onepass}}
	\label{alg}
	\textbf{Input}: $\calX$, hyperparameters $\alpha$ and $\beta$\\
	\textbf{Output}: $\calY$ and  $\calW$
	\begin{algorithmic}[1] 
		\STATE Initialize qualities $\calW = \{w_i | i \in [M]\}$.
		\FOR{$t \in [T]$}
		\STATE Estimate true label $\hat{y}_t$ by Equation (\ref{eq_yt});
		\FOR{$i \in [M]$}
		\STATE Update worker quality $\hat{w}_i^t$ by Equation (\ref{eq_wit});
		\ENDFOR
		\ENDFOR
		\STATE \textbf{return} $\calY = \{\hat{y}_t | t \in [T]\}$, $\calW = \{\hat{w}_i^T | i \in [M]\}$.
	\end{algorithmic}
\end{algorithm}
\section{Analyses}
In this section, we first theoretically prove that (a) the estimated worker quality given in Equation (\ref{eq_wit}) converges, and (b) the error of estimated worker quality can be bounded. Then we analyze the space and time complexities of LA\textsuperscript{onepass}, and compare them with that of the existing iterative algorithms and MV.

\subsection{Convergence of Estimated Worker Quality}
We analyze the convergence of worker quality in Equation (\ref{eq_wit}) under the assumption that most workers are honest and they do not incorrectly label tasks on purpose. This assumption was verified empirically \cite{yuen2011survey,gadiraju2015understanding}, and is also supported by our experimental results showing that the mean accuracy of MV is over 80\%. This assumption ensures Equation (\ref{eq_yt}) can accurately estimate true labels \cite{karger2014budget}. With this in mind, we can present the convergence of $w_i$ in the following theorem.

\begin{theorem} \label{theorem_1}
	Let $f_t(\calW)$ be the joint posterior probability of worker qualities at time-slice $t$, and $L_t(\calW) \equiv \log f_t(\calW)$:
	\begin{align} \label{eq_biglt}
		\begin{split}
			L_t(\calW) = \sum_{i=1}^M  & (C_{i,t} + \alpha-1) \log w_i^t \\
			& + (t-C_{i,t} + \beta - 1) \log(1-w_i^t)
		\end{split}		
	\end{align}
	then $\calW = \{\hat{w}_i^t | i \in [M]\}$ in Equation (\ref{eq_wit}) converges to the minimizer $\calW_t^* = \arg\min_{\calW}L_t(\calW)$ at rate of $o(1/\sqrt{t})$.
\end{theorem}

The proof of Theorem \ref{theorem_1} is in Appendix A. This theorem shows that worker quality estimated by Equation (\ref{eq_wit}) converges as fast as $o(1/\sqrt{t})$ even if it traverses all the labels only once. Moreover, the error of worker quality due to estimation can be bounded by the following corollary.

\begin{corollary} \label{corollary_1}
	$
	p(|\calW - \calW_t^*|\le\epsilon /\sqrt{t}) \ge \Phi(\epsilon) - \Phi(-\epsilon)
	$
	where $\epsilon$ is a positive real value, and $\Phi(\cdot)$ is the CDF of standard Normal distribution.
\end{corollary}
The proof of Corollary \ref{corollary_1} is in Appendix B. Corollary \ref{corollary_1} states that the error of estimated worker quality can be bounded tighter with high probability as $t$ grows. We will also empirically verify it in the experiment.

\subsection{Space and Time Complexity}
We analyze and compare the space and time complexities of LA\textsuperscript{onepass},  the existing iterative algorithms and MV. The comparison is summarized in Table \ref{tbl_sc_tc}.

\begin{table}[]
	\centering
	\begin{tabular}{cccc}
		& MV  & LA\textsuperscript{onepass} & \begin{tabular}[c]{@{}c@{}} Iterative framework\\ Algorithm \ref{alg1}\end{tabular} \\ \hline
		Space &     $\mathcal{O}(M+T)$       &    $\mathcal{O}(M+T)$              &  $\mathcal{O}(TM)$  \\ \hline
		Time & $\mathcal{O}(TM)$           &       $\mathcal{O}(TM)$           & $\mathcal{O}(ITM)$ \\ \hline
	\end{tabular}
	\caption{Space and time complexities comparison \label{tbl_sc_tc}} 
	
\end{table}
\subsubsection{Space complexity (SC). }In Algorithm \ref{alg}, LA\textsuperscript{onepass} needs to initialize $\calW$ for all the workers with $\mathcal{O}(M)$ space. It also needs to reserve $\mathcal{O}(T)$ space for storing the estimated true labels. The SC of caching hyperparameters is $\mathcal{O}(1)$. Additionally, the algorithm needs to maintain $C_{i,t}$ and $t$ for each worker with $\mathcal{O}(M)$ space. At each $t$, it needs to load the labels of task $t$, which is at most $M$. After the true label of task $t$ is estimated, the labels of task $t$ can be discarded. Therefore, the SC of loading/storing labels is $\mathcal{O}(M)$. The overall SC of LA\textsuperscript{onepass} is $\mathcal{O}(M+T)$.

As to Algorithm \ref{alg1}, the SC of caching worker qualities and true labels are the same as LA\textsuperscript{onepass}. However, it needs to load the entire dataset with $\mathcal{O}(TM)$ space. The overall SC of Algorithm \ref{alg1} is $\mathcal{O}(TM)$.

As to MV, it needs to reserve a space for storing the estimated true labels with $\mathcal{O}(T)$ space. For each task, it needs to load the related labels with $\mathcal{O}(M)$ space. The overall SC of MV is $\mathcal{O}(M+T)$. 
\subsubsection{Time complexity (TC). } As shown in Algorithm \ref{alg}, LA\textsuperscript{onepass} estimates one true label and updates all the worker qualities at every time-slice. Estimating one true label aggregates at most $M$ labels, whose TC is $\mathcal{O}(M)$. Updating one worker quality takes $\mathcal{O}(1)$ as in Equation (\ref{eq_wit}). In total it takes $\mathcal{O}(M)$ to update all the worker qualities at one time-slice. Given total $T$ tasks, the overall TC of LA\textsuperscript{onepass} is $\mathcal{O}(TM)$.

Algorithm \ref{alg1} needs to update all the estimated true labels and worker qualities in each iteration, and the TC in one such iteration is $\mathcal{O}(TM)$, which equals to the TC of LA\textsuperscript{onepass}. Assume the algorithms take $I$ iterations to converge, the overall TC of Algorithm \ref{alg1} is $\mathcal{O}(ITM)$.

MV estimates one true label by aggregating at most $M$ labels with $\mathcal{O}(M)$ time. There are $T$ tasks in total, so the overall TC is $\mathcal{O}(TM)$.

\section{Extensions} \label{sec_ext}
In this section, we present two extensions. The first extension can improve the accuracy of LA\textsuperscript{onepass} by traversing the crowdsourced labels again. The second extension describes how the proposed algorithms aggregate labels online.

\subsection{Two Pass Algorithm}
LA\textsuperscript{onepass} in Algorithm \ref{alg} estimates the true labels one after another. It traverses all the labels only once. As the algorithm aggregates more labels, the estimated worker qualities will converge, and the estimated true labels will be more accurate. However, this also raises one problem: the true labels estimated early in the process may not be accurate because the worker qualities were yet to converge. To solve this problem, we propose LA\textsuperscript{twopass}, a simple extension of LA\textsuperscript{onepass}, that traverses the labels twice. LA\textsuperscript{twopass} uses the converged worker qualities estimated by LA\textsuperscript{onepass} to perform weighted majority voting (WMV) by traversing all the labels once again to re-estimate true labels. We choose to use WMV as in Equation (\ref{eq_yt2}). It is shown that this WMV rule has provable theoretical guarantee \cite{li2014error}. LA\textsuperscript{twopass} adds little overhead over LA\textsuperscript{onepass} because it does not estimate worker qualities in the second pass. Therefore, the second pass of LA\textsuperscript{twopass} can be performed as efficient as MV, and it also has the same SC and TC as that of LA\textsuperscript{onepass}.
\begin{align}
	v_i &= K \hat{w}_i - 1, \hat{y}_t = \arg\max_k \{ \sum_{i=1}^M v_i \mathbbm{1}(x_{i,t} = k) | k \in [K]  \} \label{eq_yt2}
\end{align}
\subsection{Online Aggregation}
Since we model LA as a dynamic system, LA\textsuperscript{onepass} and LA\textsuperscript{twopass} can naturally be configured to aggregate labels online. Specifically, assuming labels are arrived sequentially in chunks. Each chunk contains labels about a few tasks, and the proposed algorithms can handle the extreme case where each chunk only contains one task's labels.  LA\textsuperscript{onepass} estimates worker qualities and true labels for tasks in the present chunk. LA\textsuperscript{twopass} uses worker qualities estimated by LA\textsuperscript{onepass} and re-estimates true labels in this chunk. After the true labels in this chunk are estimated, the labels in this chunk can be discarded. The information about worker qualities is retained in the posterior worker quality distributions for estimating the true labels of tasks in the next chunk. Therefore, LA\textsuperscript{onepass} and LA\textsuperscript{twopass} have the ability to aggregate labels online without re-visiting historical labels.

\section{Experiments}
In this section, we present the experimental results to evaluate LA\textsuperscript{onepass} and LA\textsuperscript{twopass}. Additional results with experimental environment and dataset properties can be seen in Appendix C. The code of our algorithms is also submitted as supplementary file.
\subsection{Setup}

\subsubsection{Datasets. }We use 20 publicly-available real-world datasets for evaluation. They are collected from 4 sources \cite{josephy2014workshops,zhang2014spectral,venanzi2015activecrowdtoolkit,zheng2017truth}, covering a wide range of tasks such as sentiment analysis, entity resolution and face recognition. The sizes of the datasets also vary from 1,720 to 569,282. The statistics of the datasets are summarized in Table \ref{tbl_datasets}, and their sources can be seen in Appendix C.2. Note the last column of Table \ref{tbl_datasets} shows the number of ground-truth labels available in each dataset. The ground-truth labels are only used for evaluation, but not used as input to LA algorithms.

%
	%

\begin{table}
	\centering
	\begin{tabular}{cccccc}
		\hline Dataset & $T$ &  $M$ & $K$ & \#label & \#truth \\\hline
		senti & 98980 & 1960 & 5 & 569282 & 1000  \\
		fact & 42624 & 57 & 3 & 214960 & 576 \\
		CF & 300 & 461 & 5 & 1720 & 300  \\
		CF\_amt & 300 & 110 & 5 & 6025 & 300 \\
		MS & 700 & 44 & 10 & 2945 & 700 \\
		dog & 807 & 109 & 4 & 8070 & 807  \\
		face & 584 & 27 & 4 & 5242 & 584 \\
		adult & 11040 & 825 & 4 & 89948 & 333 \\
		
		web & 2665 & 177 & 5 & 15567 & 2653  \\
		SP & 4999 & 203 & 2 & 27746 & 4999 \\
		SP\_amt & 500 & 143 & 2 & 10000 & 500 \\
		ZC\_all & 2040 & 78 & 2 & 20372 & 2040 \\
		ZC\_in & 2040 & 25 & 2 & 10626 & 2040  \\
		ZC\_us & 2040 & 74 & 2 & 11271 & 2040  \\
		prod & 8315 & 176 & 2 & 24945 & 8315 \\
		tweet & 1000 & 85 & 2 & 20000 & 1000 \\
		bird & 108 & 39 & 2 & 4212 & 108 \\
		trec & 19033 & 762 & 2 & 88385 & 2275 \\
		rte & 800 & 164 & 2 & 8000 & 800 \\
		smile & 2134 & 64 & 2 & 30319 & 159 \\
		\hline
	\end{tabular}
	\caption{Statistics about the datasets. \label{tbl_datasets}}
	
\end{table}

%


\subsubsection{Methods. } The compared methods include MV, DS \cite{dawid1979maximum}, LFC \cite{raykar2010learning}, EBCC \cite{li2019exploiting}, BiLA \cite{hong2021online}, ZC \cite{demartini2012zencrowd}, IWWV \cite{li2014error} and LAA \cite{li2017aggregating}. Descriptions of them can be seen in Prior Work sub-section.

\subsubsection{Metrics. }We use \textit{accuracy} to evaluate the effectiveness of an algorithm. The accuracy is defined as the ratio between the number of correctly estimated true labels and number of tasks. The efficiency is evaluated by $\lg(sec)$, which is defined as the base 10 logarithm of an algorithm's runtime in seconds. If $\lg(sec)$ is increased by $1$, the algorithm's runtime is increased 10 times.

\subsubsection{Hyperparameters.} The proposed algorithms require to set two hyperparameters $\alpha$ and $\beta$. Usually the iterative LA methods select hyperparameters by consulting the results of MV. However, running MV requires to traverse the entire dataset once, but the proposed algorithms only traverse all the labels at most twice. They cannot afford to run MV for initialization. Therefore, we set $\alpha = 2$ and $\beta = 2$ for all datasets. Later we will show that the proposed algorithms are insensitive and robust to hyperparameters.

\subsubsection{Order of tasks. }The proposed algorithms estimate true labels sequentially. In order to show they are robust to the order of tasks, we run the proposed algorithms 10 times with shuffled tasks over each dataset, and the accuracies of the proposed algorithms are averaged over the 10 runs for each dataset. We also use this strategy in the online experiment.

\subsection{Offline Experimental Results}

We first present the experimental results conducted offline when all the labels are fed into the algorithm at once. The mean accuracies and $\lg(sec)$ of each method over 20 datasets are summarized in Figure \ref{fig_acc_time_offline}. The detailed per-dataset results can be seen in Appendix C.3. From Figure \ref{fig_acc_time_offline}, first we can observe that LA\textsuperscript{twopass} and LA\textsuperscript{onepass} rank second and fourth over all the methods in terms of accuracy. This demonstrates that the proposed algorithms can effectively estimate true labels even if they traverse the entire labels at most twice. Second, LA\textsuperscript{twopass} is more accurate than LA\textsuperscript{onepass} on average, which shows that traversing the labels again indeed improves accuracy. Third, the mean accuracy of LA\textsuperscript{twopass} is only second to EBCC. To the best of authors' knowledge, EBCC is the best performing LA method in general because it learns worker correlation while estimating true labels. However, EBCC is extremely inefficient, it is three orders of magnitude slower than the proposed algorithms, making it unscalable to large datasets. For example, we find that EBCC takes about 4.5 hours to aggregate \textit{senti} while LA\textsuperscript{twopass} only takes about 5 seconds. Lastly, the proposed algorithms also run much faster than the methods that model worker qualities. It is also worth noting that the runtime of LA\textsuperscript{onepass} is very close to that of MV, but is much more accurate than MV. 
\begin{figure}[!t]
	\centering
	\includegraphics[width=\columnwidth]{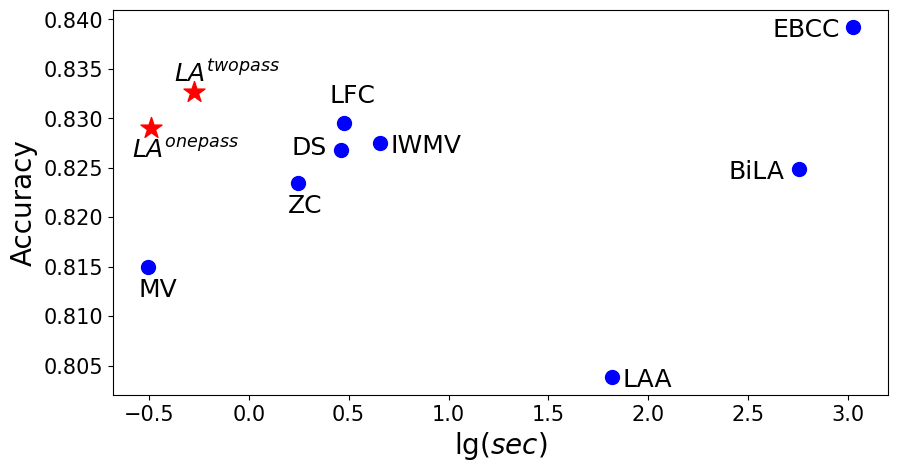}
	\caption{Offline experimental results \label{fig_acc_time_offline}}
\end{figure}

\subsubsection{Comparison to MV.} We perform one-sided Wilcoxon signed rank test \cite{wilcoxon1992individual} on every method against MV based on each method's accuracies on 20 datasets, and the results are summarized in Table \ref{tbl_mean_acc}. The results of Wilcoxon test tell whether each method is significantly more accurate than MV and at what level. We use two significance levels whose p-value thresholds are 0.01(**) and 0.05(*), respectively, and \textit{W\_} denotes the statistic of a method summing up the ranks of datasets that this method is compared with MV. From Table \ref{tbl_mean_acc}, we can see only EBCC, LA\textsuperscript{twopass} and LA\textsuperscript{onepass} are at ** significance level. This provides a very strong evidence that LA\textsuperscript{twopass} and LA\textsuperscript{onepass} perform significantly better than MV in terms of accuracy. 

\begin{table}[htp]
	\centering
	\begin{tabular}{cccc}
		\hline
		Method         & W\_ & significance level & p-value \\ \hline
		
		DS                                                               & 134   &                                                      & 0.1471  \\
		LFC                                                             & 136   &                                                      & 0.1305  \\
		IWMV                                                           & 165   & *                                                    & 0.012   \\
		EBCC                                                            & 182   & **                                                   & 0.0014  \\
		LAA                                                             & 80    &                                                      & 0.8256  \\
		ZC                                                               & 124   &                                                      & 0.2490  \\
		BiLA                                                  & 147   &                                                      & 0.0615  \\\hline
		LA\textsuperscript{twopass}                                                       & 178   & **                                                   & 0.0024  \\
		LA\textsuperscript{onepass}                                                     & 176   & **                                                   & 0.0032 \\\hline
	\end{tabular}
	\caption{One-sided Wilcoxon signed rank test results \label{tbl_mean_acc}}
\end{table}

\subsubsection{Hyperparameter robustness. } 
\begin{figure}[!t]
	\includegraphics[width=\columnwidth]{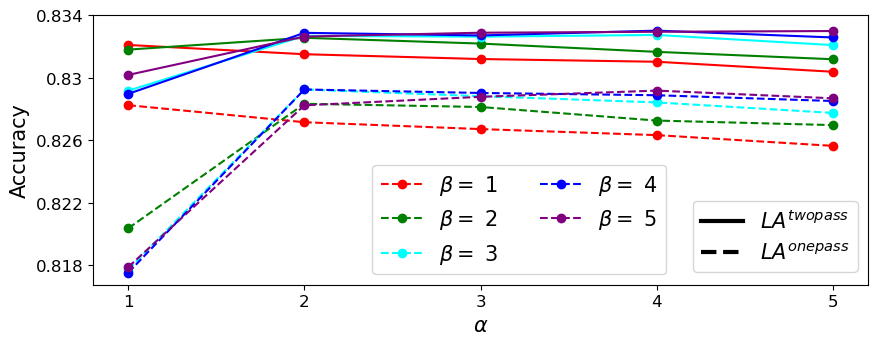}
	\caption{Accuracies of the proposed algorithms under different initial settings. \label{fig_hyper}}
\end{figure}

We report the accuracies of the proposed algorithms initialized by different parameters in Figure \ref{fig_hyper}. We vary $\alpha$ and $\beta$ from $(1,1)$ to $(5,5)$ to generate 25 different initial settings.  From Figure \ref{fig_hyper}, we can see that the gaps between the best and worst accuracies of LA\textsuperscript{twopass} and LA\textsuperscript{onepass} are 0.005 and 0.1, respectively. This shows the proposed algorithms are robust to hyperparameters. Note that the LA\textsuperscript{onepass}'s accuracies are relatively low when  $\alpha=1$ and $\beta>1$. This is because the initial worker qualities under these combinations are very low, which take time for the worker qualities to converge. But we can see that even in these cases LA\textsuperscript{onepass} can eventually estimate worker qualities accurately because LA\textsuperscript{twopass} has higher accuracies with the same initialization. The result shows that the proposed algorithms are robust to hyperparameters, which differs from some iterative algorithms requiring delicate hyperparameters. For example, EBCC requires to carefully set 6 hyparameters in order to achieve the reported accuracies.


\subsubsection{Worker Quality Convergence Study.} 
\begin{figure}
	\begin{subfigure}{0.99\columnwidth}
		\includegraphics[width=\columnwidth]{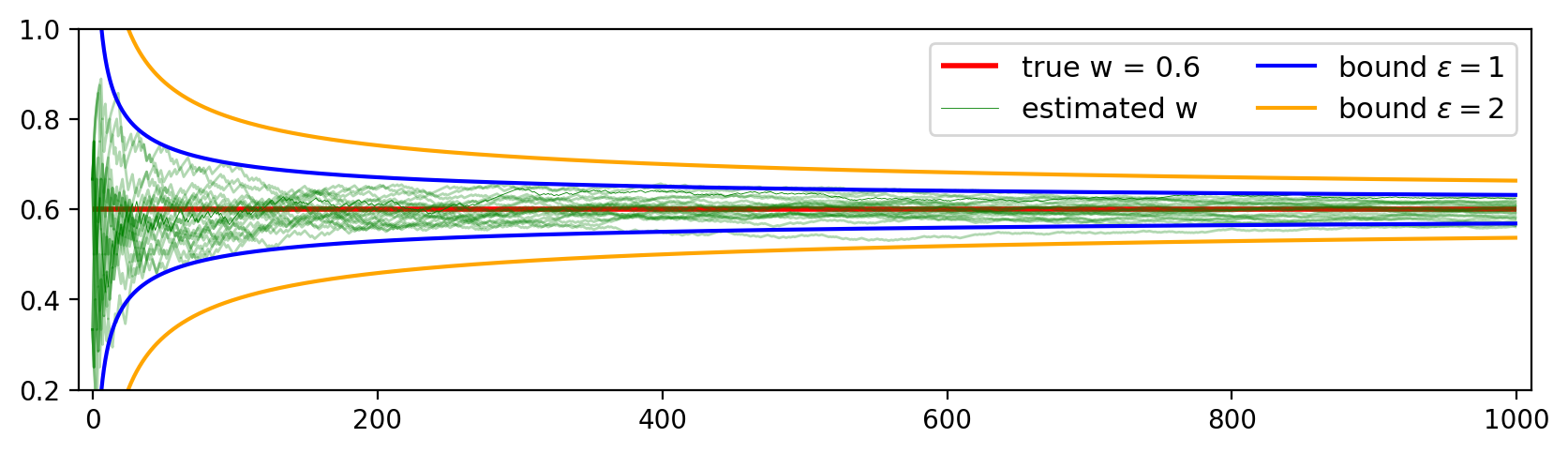}
	\end{subfigure}
	
	\begin{subfigure}{0.99\columnwidth}
		\includegraphics[width=\columnwidth]{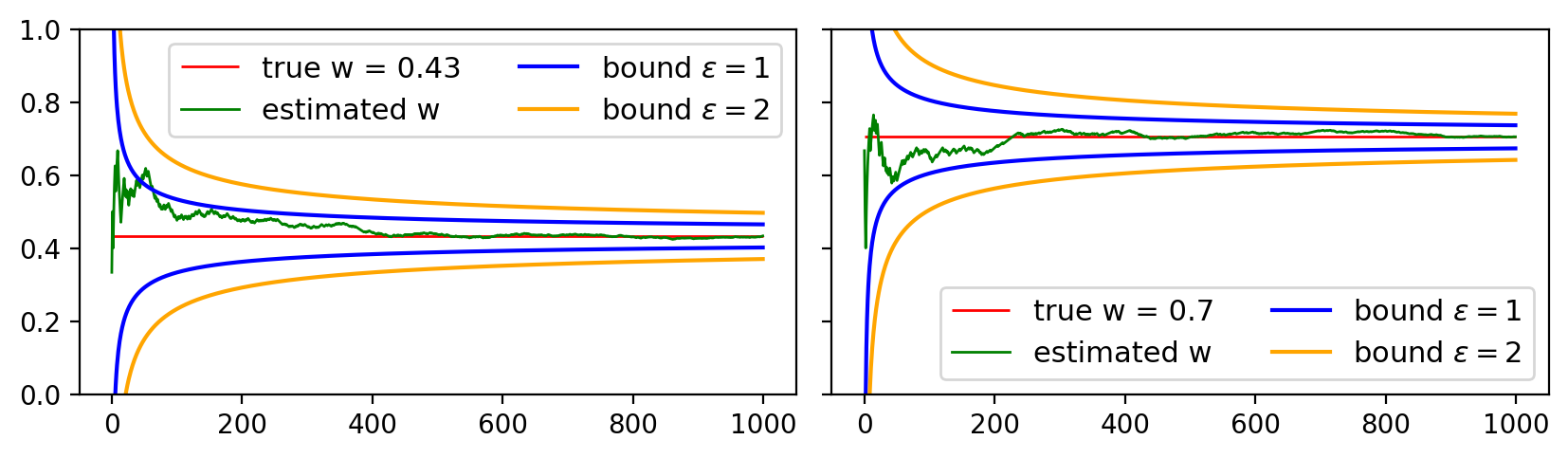}
	\end{subfigure}
	\caption{Simulated experimental results. \label{fig_simulation}}
\end{figure}

We perform two simulations to verify the theoretical convergence and error bound of LA\textsuperscript{onepass}. In the first simulation, we generate 20 workers with the same true worker quality $0.6$. These workers label 1000 tasks with 4 classes, and their labels are generated according to their worker qualities. Then the generated labels are fed to LA\textsuperscript{onepass} to estimate worker qualities. The results are illustrated in Figure \ref{fig_simulation} (top). In Figure \ref{fig_simulation} (top), the red line shows the true worker quality. The green traces show the evolution of estimated worker qualities. The blue and orange curves are the error bounds computed by taking $\epsilon = \{1,2\}$ in Corollary \ref{corollary_1}, which bound the error with at least $67\%$ and $95\%$ probabilities, respectively. In the second simulation, we also generate 20 workers but their worker qualities are sampled from $[0.4,0.7]$, the other settings remain the same as the first one. Due to space limitation, we randomly select two workers and present their estimated worker qualities in Figure \ref{fig_simulation} (bottom). The complete results of the second simulation are in Appendix C.4. From Figure \ref{fig_simulation}, we can observe that the estimated worker qualities can converge to their true qualities regardless of whether the workers have the same quality or not, and the errors due to estimation can also be bounded with high probability. The results empirically confirm the claims in Theorem \ref{theorem_1} and Corollary \ref{corollary_1}.


\subsection{Online Experimental Results}
\begin{figure}[!ht]
	\includegraphics[width=\columnwidth]{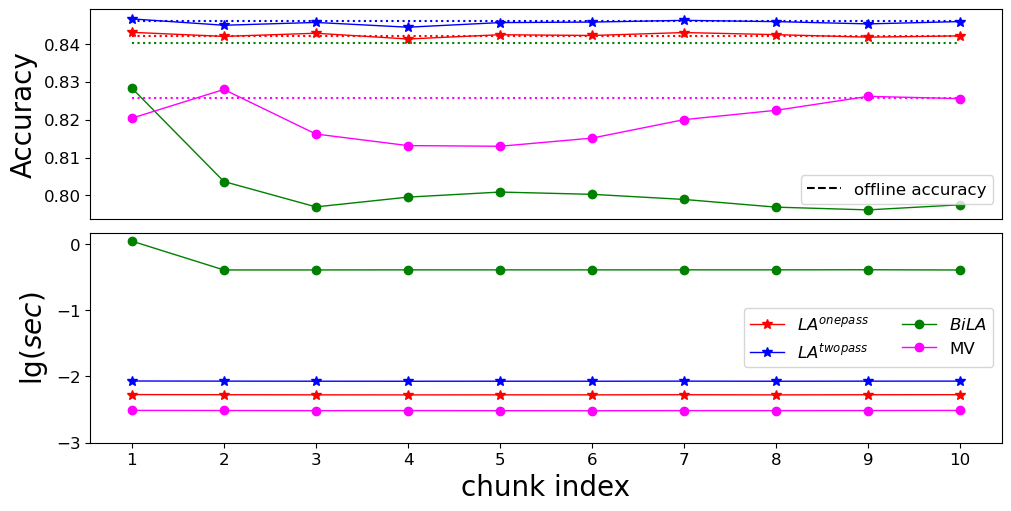}
	\caption{Online experimental results. \label{fig_online}}
\end{figure}
Finally, we perform experiments to show that the proposed algorithms can aggregate labels online over 15 selected datasets (listed in Appendix C.5) where each has at least 95\% tasks with ground-truths for evaluation. We divide the tasks in each dataset evenly into 10 chunks, and feed their labels in each chunk sequentially to the algorithms.  We compare the proposed algorithms with MV and BiLA \cite{hong2021online}, which can aggregate labels online.


We report the averaged accuracy and $\lg(sec)$ in Figure \ref{fig_online}. The detailed per-dataset results are in Appendix C.5. From the figure, we can observe that the proposed algorithms can accurately estimate true labels online, and the accuracies of the proposed algorithms and MV converge to their offline counterparts. However, BiLA does not converge to its offline  accuracy. This is because BiLA uses neural network in its model, which requires a large number of labels to train in order to achieve a decent performance. However, there are inadequate labels for BiLA to train online in the datasets. We also observe that the proposed algorithms can efficiently aggregate labels online, whose runtimes are close to that of MV. In contrast, although BiLA can aggregate labels online without re-visiting historical data, it needs to traverse the labels in the present chunk multiple times to train its model, which cannot aggregate labels in real-time.

\section{Conclusion}
This paper presents a novel light-weight method for aggregating crowdsourced labels. We treat LA as a dynamic system and model it by a Dynamic Bayesian network. Two algorithms, LA\textsuperscript{twopass} and LA\textsuperscript{onepass}, are derived from the model, which can aggregate labels by traversing the dataset at most twice. We prove the worker quality estimated by LA\textsuperscript{onepass} converges as fast as $o(1/\sqrt{t})$, its error can be bounded, and show the space and time complexities of the proposed algorithms are equal to those of MV. Experiments demonstrate that the proposed algorithms can effectively and efficiently aggregate labels offline and online.

\bibliographystyle{unsrt}  
\bibliography{references}  

\appendix

\section{Appendix A. Proof of Theorem \ref{theorem_1} \label{appendix_proof}}
In order to prove the theorem, we first present Lemma \ref{lemma1}.
\begin{lemma} \label{lemma1}
	Let $\{f_t(\calW), t = 1,2,\dots\}$ be a sequence of posterior probability density functions $p(\calW|C_t)$ of random vectors defined on $[0,1]^M$. Define $L_t(\calW) \equiv \log f_t(\calW)$ as in Equation (\ref{eq_biglt}). Suppose for each $t$, there exists a strict local maximum, $\calW_t^*$, of $L_t(\calW)$.  Then the posterior distribution $p(\calW|C_t)$ satisfies asymptotic normality:
	\begin{align} \label{eq_asy_normal}
		(-\nabla^2L_t(\calW_t^*))^{1/2} (\calW - \calW_t^*) \xrightarrow{d} N(0,1) \text{ as } t \rightarrow \infty,
	\end{align}
	where $C_t = \sum_{i=1}^M C_{i,t}$ is the number of correctly labeled tasks by all workers up to time-slice $t$.
\end{lemma}

\begin{proof}
	We use Theorem 2.1 in \cite{chen1985asymptotic} to prove this lemma. Theorem 2.1 in \cite{chen1985asymptotic} states if the following conditions (P1-2 and C1-3) are satisfied, then the asymptotic normality property in Equation (\ref{eq_asy_normal}) holds.
	
	P1. $\nabla \log p(\calW_t^*|C_t)  = 0$.
	
	P2. $\Sigma_t \equiv \{ -\nabla^2 \log p(\calW_t^*|C_t)  \}^{-1}$ is positive definite.
	
	C1. ``Steepness'': as $t \rightarrow \infty$, $\sigma_t^2 \rightarrow 0$ where $\sigma_t^2$ is the largest eigenvalue of $\Sigma_t$.
	
	C2. ``Smoothness'':  for any $\epsilon > 0$, there exists an integer $N$ and $\delta > 0$ such that, for any $t > N$, and $\calW' \in H(\calW_t^*;\delta) = \{|\calW' - \calW_t^*| < \delta\}$, $\nabla^2 \log p (\calW'|C_t)|$ satisfies $I - A(\epsilon) \le \nabla^2 \log p(\calW'|C_t)|\{ \nabla^2 \log p(\calW_t^*|C_t)| \}^{-1} \le I + A(\epsilon)$, where $I$ denotes the identity matrix with an appropriate size and $A(\epsilon)$ is the positive semi-definite symmetric matrix with the largest eigenvalue goes to $0$ as $\epsilon \rightarrow 0$.
	
	C3. ``Concentration'': for any $\delta > 0$, $\int_{H(\calW;\delta)} p(\calW|C_t) d\calW \rightarrow 1$ as $t \rightarrow \infty$.
	
	In the rest of the proof, we will show the satisfactions of these conditions.
	
	\textbf{Proof of P1 and P2}. Since $\calW_t^*$ is a local maximum of $L_t$, the satisfaction of P1 is straightforward. The Hessian of $L_t$ is
	\begin{align} \label{eq_hessian}
		\begin{split}
			\nabla^2 L_t (\calW_t^*) = diag(-\frac{C_{i,t} + \alpha - 1}{(w_i^*)^2} -\frac{t-C_{i,t} + \beta - 1}{(1-w_i^*)^2})_i
		\end{split}
	\end{align}
	where $C_{i,t}$ is the number of correctly labeled tasks by worker $i$ up to $t$. It can be observed that $\nabla^2 L_t (\calW_t^*)$ is negative definite because $\nabla^2 L_t (\calW_t^*)$ is a diagonal matrix whose diagonal entries are negative given reasonable and small hyperparameters $\alpha$ and $\beta$. Therefore $\Sigma_t$ is positive definite and P2 satisfies.
	
	\textbf{Proof of C1}. As $t \rightarrow \infty$, the diagonal entries of $\nabla^2 L_t (\calW_t^*)$ approach $-\infty$. Hence the diagonal entries of $\Sigma_t$ approaches $0$. It implies all the eigenvalues of $\Sigma_t$ go to $0$ as $t \rightarrow \infty$. Therefore, C1 is satisfied.
	
	\textbf{Proof of C2}. C2 is straightforward because all the entries in $\nabla^2 L_t(\calW)$ are continuous with respect to each $w_i$ in its domain.
	
	\textbf{Proof of C3}. By setting $\nabla L_t(\calW) = 0$, we can easily find $(\calW_t^*)_i$ has the form as given in Equation (\ref{eq_wit}), which is the mode of a posterior distribution $Beta(C_{i,t} + \alpha,t-C_{i,t} + \beta)$. The variance of the posterior distribution is  $\frac{ (C_{i,t}+\alpha) (t-C_{i,t} + \beta)}{(t+\alpha+\beta)^2 (t+\alpha+\beta+1)}$. Because $C_{i,t} \le t$, the denominator of the variance dominates the numerator. Therefore the variance approaches $0$ as $t \rightarrow \infty$. This means $E_{p(\calW|C_t)}[\calW - \calW_t^*] \rightarrow 0$. Therefore C3 satisfies.
\end{proof}

The lemma shows the posterior distribution of worker quality converges as $t \rightarrow \infty$, and it converges to the minimizer $\calW_t^*$ of $L_t(\calW)$. 

From Lemma \ref{lemma1} we can take the expectation on the asymptotic distribution in Equation (\ref{eq_asy_normal}) and get
\begin{align}
	E[	(-\nabla^2L_t(\calW_t^*))^{1/2} (\calW - \calW_t^*)] \rightarrow 0.
\end{align}
It implies 
\begin{align}
	|E_{p(\calW|C_t)}(\calW) - \calW_t^*| = o(1) |(-\nabla^2L_t(\calW_t^*))^{-1/2}|,
\end{align}
where $E_{p(\calW|C_t)}(\calW)$ is the posterior mean of $\calW$ at time-slice $t$. From Equation (\ref{eq_hessian}), we can see $-\nabla^2 L_t (\calW_t^*) = \Theta(t) $. Therefore, $|E_{p(\calW|C_t)}(\calW) - \calW_t^*| = o(1/\sqrt{t})$.

Moreover, the $\hat{w}_i^t$ given by Equation (\ref{eq_wit})  is the mode of posterior distribution $p(w_i^t | C_t)$. Therefore we have
\begin{align} \label{eq_mode_mean}
	\begin{split}
		(|\hat{\calW} - E_{p(\calW|C_t)}(\calW)|)_i &= |\hat{w}_i^t - E_{p(w_i | C_{i,t})}(w_i)| 		
	\end{split}
\end{align}
Denote the two parameters of the posterior distribution $p(w_i|C_{i,t})$ as $\alpha_{i} = C_{i,t} + \alpha$ and $\beta_{i} = t-C_{i,t} + \beta$. The mode and mean of the posterior can be written as $\frac{\alpha_{i} - 1}{\alpha_{i} + \beta_{i} - 2}$ and $\frac{\alpha_{i}}{\alpha_{i} + \beta_{i}}$, respectively. Therefore, we can write out Equation (\ref{eq_mode_mean}) as
\begin{align}
	\begin{split}
		&(|\hat{\calW} - E_{p(\calW|C_t)}(\calW)|)_i = |\frac{\alpha_{i} - 1}{\alpha_{i} + \beta_{i} - 2} - \frac{\alpha_{i}}{\alpha_{i} + \beta_{i}}| \\
		& = |\frac{(\alpha_{i} - 1)(\alpha_{i} + \beta_{i}) - \alpha_{i}(\alpha_{i} + \beta_{i} - 2)}{(\alpha_{i} + \beta_{i} - 2) (\alpha_{i} + \beta_{i})}| \\
		& =| \frac{(\alpha_{i} - \beta_{i})}{(\alpha_{i} + \beta_{i} - 2) (\alpha_{i} + \beta_{i})}| \\
		&\le | \frac{1}{(\alpha_{i} + \beta_{i} - 2)}| = |\frac{1}{t+\alpha+\beta - 2}| = \Theta(1/t) = o(1/\sqrt{t})
	\end{split}
\end{align}
Hence, $|\hat{\calW} - E_{p(\calW|C_t)}(\calW)| = o(1/\sqrt{t})$. By triangle inequality, we have
\begin{align}
	|\hat{\calW} - \calW_t^*| \le |\hat{\calW} - E_{p(\calW|C_t)}(\calW)| + |E_{p(\calW|C_t)}(\calW) - \calW_t^*|.
\end{align}
We have shown that $|E_{p(\calW|C_t)}(\calW) - \calW_t^*| = o(1/\sqrt{t})$ and $|\hat{\calW} - E_{p(\calW|C_t)}(\calW)| = o(1/\sqrt{t})$, so $|\hat{\calW} - \calW_t^*| = o(1/\sqrt{t})$, which proves the theorem.

\section{Appendix B. Proof of Corollary \ref{corollary_1}}
From Equation (\ref{eq_asy_normal}) and given the fact that $\calW_t^*$ is a vector of scalars, we can derive 
\begin{align} \label{eq_What}
	\hat{\calW}  \xrightarrow{d} N(\calW_t^*,\Sigma_t),
\end{align}
where $\Sigma_t \equiv \{ -\nabla^2 \log p(\calW_t^*|C_t)  \}^{-1} = \{ -\nabla^2 L_t(\calW_t^*) \}^{-1}$ as defined in Lemma \ref{lemma1}. By transformation of Normal distribution, we have
\begin{align}
	p(|\hat{\calW} - \calW_t^*| \le \epsilon \Sigma_t^{1/2})  = \Phi(\epsilon) - \Phi(-\epsilon),
\end{align}
where $\epsilon$ is a positive real number and $\Phi(\cdot)$ is the CDF of standard Normal distribution. Since $\Sigma_t^{1/2}$ is the standard deviation of the Normal distribution in Equation (\ref{eq_What}), $p(|\hat{\calW} - \calW_t^*| \le \epsilon \Sigma_t^{1/2})$ can be interpreted as \textit{the probability that $\hat{\calW}$ falls within $\epsilon$ standard deviations away from $\calW_t^*$}.

From Equation (\ref{eq_hessian}), we have $\big(\nabla^2 L_t (\calW_t^*)\big)_i \le -t$, which implies that $(\Sigma_t)_i^{1/2} \le 1/\sqrt{t}$. Therefore, 
\begin{align}
	p(|\hat{\calW} - \calW_t^*| \le \epsilon \Sigma_t^{1/2}) \le p(|\hat{\calW} - \calW_t^*| \le \epsilon/\sqrt{t}),
\end{align}
which implies
\begin{align}
	p(|\hat{\calW} - \calW_t^*| \le \epsilon/\sqrt{t}) \ge \Phi(\epsilon) - \Phi(-\epsilon).
\end{align}

\section{Appendix C. Experiments}
In this section, we provide additional information about the experiments.

\subsection{C.1. Implementation and Experiment Environment}
We run all experiments on a Ubuntu desktop with an AMD 5900 CPU and 16 GB memory. 

The algorithms in the experiments are all implemented in Python. We implement MV, IWMV and the proposed LA\textsuperscript{twopass} and LA\textsuperscript{onepass} by ourselves. The implementations of DS, LFC and ZC along with the hyperparameters are from the Truth Inference Project \cite{zheng2017truth}. The implementations of EBCC \cite{li2019exploiting}, LAA \cite{li2017aggregating} and BiLA \cite{hong2021online} along with the hyperparameters are from the papers' authors.

\subsection{C.2. Datasets and Pre-processing}
The links and sources of the used datasets are listed below. The descriptions of the datasets can also be found in the given links.
\begin{itemize}
	\item \textit{senti} and \textit{fact} are collected from  CrowdScale 2013 \cite{josephy2014workshops}. They can be downloaded at \url{https://sites.google.com/site/crowdscale2013/home}.
	\item \textit{MS}, \textit{ZC\_in}, \textit{ZC\_us}, \textit{ZC\_all}, \textit{SP}, \textit{SP\_amt}, \textit{CF} and \textit{CF\_amt} are from Active Crowd Toolkit project \cite{venanzi2015activecrowdtoolkit}. They can be downloaded at \url{https://github.com/orchidproject/active-crowd-toolkit}.
	\item \textit{prod}, \textit{tweet}, \textit{dog}, \textit{face}, \textit{smile} and \textit{adult} are collected from Truth Inference Project \cite{zheng2017truth}. They can be downloaded at \url{http://dbgroup.cs.tsinghua.edu.cn/ligl/crowddata/}.
	\item \textit{bird}, \textit{rte}, \textit{web} and \textit{trec} are used in \cite{zhang2014spectral}. They can be downloaded at \url{https://github.com/zhangyuc/SpectralMethodsMeetEM}.
\end{itemize}

\subsubsection{Pre-processing.} We perform two pre-processing steps after the datasets are collected.
\begin{enumerate}
	\item  We re-index the tasks from $0$ to $T-1$, workers from $0$ to $M-1$ and classes from $0$ to $K-1$. It complies with the indexing rule of Python.
	\item We delete the ground-truth labels whose tasks do not appear in the crowdsourced labels. For example, there are 160 tasks with ground-truth labels in the \textit{smile} dataset. But we find one of the 160 tasks does not have any crowdsourced label. So we delete this ground-truth label.
\end{enumerate}
We do not modify the datasets further to keep the datasets as intact as possible.

\subsection{C.3. Offline Experiments}
The accuracies and runtimes of methods over each dataset are summarized in the following tables.

\begin{table*}[hbt!]
	\centering
	\begin{tabular}{ccccccccccc}
		\hline
		Dataset & MV & DS & LFC & IWMV & EBCC & LAA & ZC & BiLA & LA\textsuperscript{twopass} & LA\textsuperscript{onepass} \\ \hline
		senti & 0.8828  & 0.8240  & 0.818  & 0.89  & 0.879  & N/A & 0.889  & 0.779  & 0.8922  & 0.8899  \\ 
		fact & 0.9017  & 0.8507  & 0.8611  & 0.9010  & 0.8924  & 0.8958  & 0.901  & 0.941  & 0.901  & 0.901  \\ 
		CF & 0.8843  & 0.7967  & 0.8167  & 0.8833  & 0.8833  & 0.8533  & 0.88  & 0.8867  & 0.8823  & 0.8823  \\ 
		CF\_amt & 0.8558  & 0.8567  & 0.8367  & 0.8567  & 0.8533  & 0.8467  & 0.8533  & 0.8667  & 0.8564  & 0.8574  \\ 
		MS & 0.7068  & 0.7643  & 0.7743  & 0.8  & 0.7871  & 0.6957  & 0.7971  & 0.7829  & 0.7959  & 0.7951  \\ 
		face & 0.6363  & 0.6404  & 0.6404  & 0.6301  & 0.6353  & 0.6507  & 0.6284  & 0.6438  & 0.6303  & 0.6341  \\ 
		adult & 0.7577  & 0.7447  & 0.7628  & 0.7658  & 0.7478  & 0.6727  & 0.7207  & 0.7628  & 0.7655  & 0.7637  \\ 
		dog & 0.8224  & 0.8426  & 0.8426  & 0.8302  & 0.8401  & 0.8364  & 0.8302  & 0.8302  & 0.8302  & 0.8314  \\ 
		web & 0.7314  & 0.8255  & 0.8326  & 0.8424  & 0.7437  & 0.8428  & 0.8398  & 0.7791  & 0.8376  & 0.8166  \\ 
		SP & 0.8853  & 0.9148  & 0.9148  & 0.9052  & 0.9152  & 0.8794  & 0.9166  & 0.9056  & 0.9031  & 0.8949  \\
		SP\_amt & 0.9426  & 0.944  & 0.944  & 0.946  & 0.944  & 0.944  & 0.946  & 0.942  & 0.944  & 0.943  \\ 
		ZC\_all & 0.8307  & 0.7926  & 0.7922  & 0.8343  & 0.8632  & 0.7735  & 0.8299  & 0.7961  & 0.8427  & 0.8355  \\ 
		ZC\_in & 0.7402  & 0.7608  & 0.7598  & 0.7490  & 0.7721  & 0.6716  & 0.7725  & 0.7186  & 0.7463  & 0.7437  \\ 
		ZC\_us & 0.8607  & 0.8211  & 0.8211  & 0.8706  & 0.9123  & 0.8098  & 0.8578  & 0.8250  & 0.8678  & 0.8633  \\ 
		product & 0.8966  & 0.9366  & 0.9373  & 0.9274  & 0.9349  & 0.8485  & 0.9280  & 0.8946  & 0.9262  & 0.9083  \\ 
		tweet & 0.9341  & 0.9600  & 0.9600  & 0.9480  & 0.9610  & 0.9560  & 0.9510  & 0.9530  & 0.9481  & 0.9512  \\ 
		bird & 0.7593  & 0.8796  & 0.8981  & 0.7222  & 0.8611  & 0.8796  & 0.7222  & 0.8426  & 0.7537  & 0.7528  \\ 
		rte & 0.8976  & 0.9275  & 0.9275  & 0.9275  & 0.9313  & 0.9163  & 0.9250  & 0.9350  & 0.9266  & 0.9195  \\ 
		trec & 0.6521  & 0.7046  & 0.7024  & 0.5912  & 0.7037  & 0.5829  & 0.5697  & 0.6963  & 0.6315  & 0.6428  \\ 
		smile & 0.7201  & 0.7484  & 0.7484  & 0.7296  & 0.7233  & 0.7170  & 0.7107  & 0.7170  & 0.7723  & 0.7541  \\ \hline
		mean & 0.8149  & 0.8268  & 0.8295  & 0.8275  & 0.8392  & 0.8038  & 0.8235  & 0.8249  & 0.8327  & 0.8290 \\ \hline
	\end{tabular}
	\caption{Offline accuracies of methods over 20 real-world datasets.}
\end{table*}
\begin{table*}[hbt!]
	\centering
	\begin{tabular}{ccccccccccc}
		\hline
		Dataset & MV & DS & LFC & IWMV & EBCC & LAA & ZC & BiLA & LA\textsuperscript{twopass} & LA\textsuperscript{onepass} \\ \hline
		senti & 3.066  & 36.965  & 38.406  & 49.020  & 15588  & N/A & 22.077  & 10161  & 5.618  & 3.446  \\ 
		fact & 1.312  & 9.702  & 9.733  & 17.990  & 3125.148  & 42.877  & 6.182  & 137.687  & 2.027  & 1.239  \\ 
		CF & 0.010  & 0.121  & 0.171  & 0.150  & 11.042  & 14.529  & 0.055  & 7.017  & 0.017  & 0.010  \\ 
		CF\_amt & 0.010  & 0.218  & 0.229  & 0.155  & 2.213  & 2.392  & 0.124  & 2.445  & 0.020  & 0.012  \\ 
		MS & 0.022  & 0.254  & 0.238  & 0.481  & 59.165  & 6.471  & 0.136  & 4.202  & 0.053  & 0.030  \\ 
		face & 0.019  & 0.160  & 0.167  & 0.276  & 20.880  & 1.498  & 0.105  & 1.640  & 0.032  & 0.020  \\ 
		adult & 0.332  & 3.874  & 3.807  & 5.068  & 711  & 852.562  & 2.347  & 435.186  & 0.601  & 0.376  \\ 
		dog & 0.025  & 0.251  & 0.276  & 0.373  & 9.582  & 4.123  & 0.153  & 4.747  & 0.045  & 0.028  \\ 
		web & 0.083  & 0.744  & 0.709  & 1.347  & 159.522  & 30.692  & 0.388  & 30.598  & 0.156  & 0.093  \\ 
		SP & 0.158  & 0.635  & 0.621  & 1.882  & 173.864  & 10.878  & 0.391  & 23.458  & 0.225  & 0.142  \\ 
		SP\_amt & 0.016  & 0.158  & 0.190  & 0.195  & 2.258  & 2.330  & 0.099  & 2.173  & 0.034  & 0.021  \\ 
		ZC\_all & 0.064  & 0.361  & 0.362  & 0.786  & 67.306  & 3.184  & 0.250  & 4.987  & 0.094  & 0.060  \\ 
		ZC\_in & 0.070  & 0.223  & 0.230  & 0.782  & 50.154  & 2.388  & 0.132  & 2.867  & 0.093  & 0.059  \\ 
		ZC\_us & 0.065  & 0.212  & 0.237  & 0.788  & 45.786  & 3.083  & 0.155  & 5.159  & 0.097  & 0.064  \\ 
		product & 0.273  & 0.684  & 0.702  & 3.133  & 58.687  & 14.550  & 0.455  & 34.203  & 0.365  & 0.225  \\ 
		senti\_1k & 0.036  & 0.316  & 0.332  & 0.393  & 25.383  & 2.885  & 0.204  & 3.048  & 0.050  & 0.033  \\ 
		bird & 0.004  & 0.061  & 0.065  & 0.043  & 0.778  & 2.818  & 0.038  & 0.613  & 0.007  & 0.005  \\ 
		rte & 0.026  & 0.131  & 0.157  & 0.311  & 9.998  & 3.508  & 0.088  & 3.481  & 0.039  & 0.025  \\ 
		trec & 0.618  & 2.693  & 2.695  & 7.304  & 923.698  & 247.690  & 1.801  & 421.628  & 0.924  & 0.562  \\ 
		smile & 0.069  & 0.317  & 0.386  & 0.847  & 31.386  & 3.936  & 0.235  & 6.386  & 0.113  & 0.072  \\ \hline
		mean & 0.314  & 2.904  & 2.986  & 4.566  & 1053.806  & 65.915  & 1.771  & 564.626  & 0.530  & 0.326 \\ \hline
	\end{tabular}
	\caption{Offline runtime (in seconds) of methods over 20 real-world datasets.}
\end{table*}

\FloatBarrier
\subsection{C.4. Simulated Experimental Results}
\begin{figure*}[!ht]
	\centering
	\includegraphics[height=0.88\textheight,width=\textwidth]{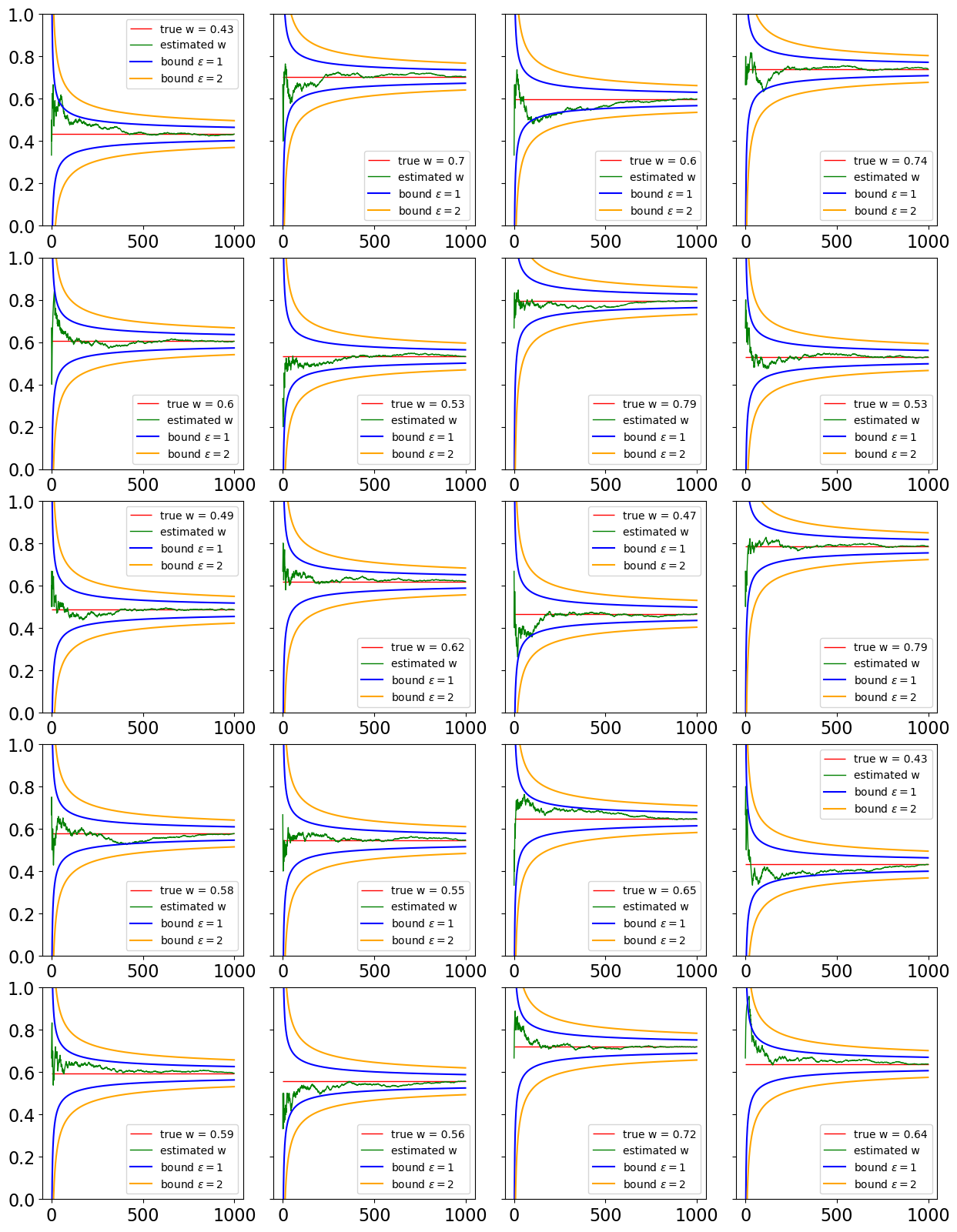}
	\caption{Complete simulated experimental results of the second simulation. x-axis is the time-slice/task index; y-axis is the worker quality.}
\end{figure*}

\FloatBarrier
\subsection{C.5. Online Experiments}
In the online experiments, we select the following 15 datasets: \textit{CF, CF\_amt, MS, dog, face, web, SP, SP\_amt, ZC\_all, ZC\_in, ZC\_us, product, tweet, bird, rte}. All the tasks in the selected datasets, except \textit{web}, have ground-truth labels for evaluation.  There are 2665 tasks in \textit{web}, and 2653 of them have ground-truth labels for evaluation. We do not select the other datasets because there are too few tasks with ground-truth labels to be distributed into each chunk for evaluation.

The accuracies and runtimes of each method performed on the selected datasets at every data chunk are illustrated in the figures on the next two pages.

\begin{figure*}[!ht]
	\centering
	\includegraphics[width=\textwidth]{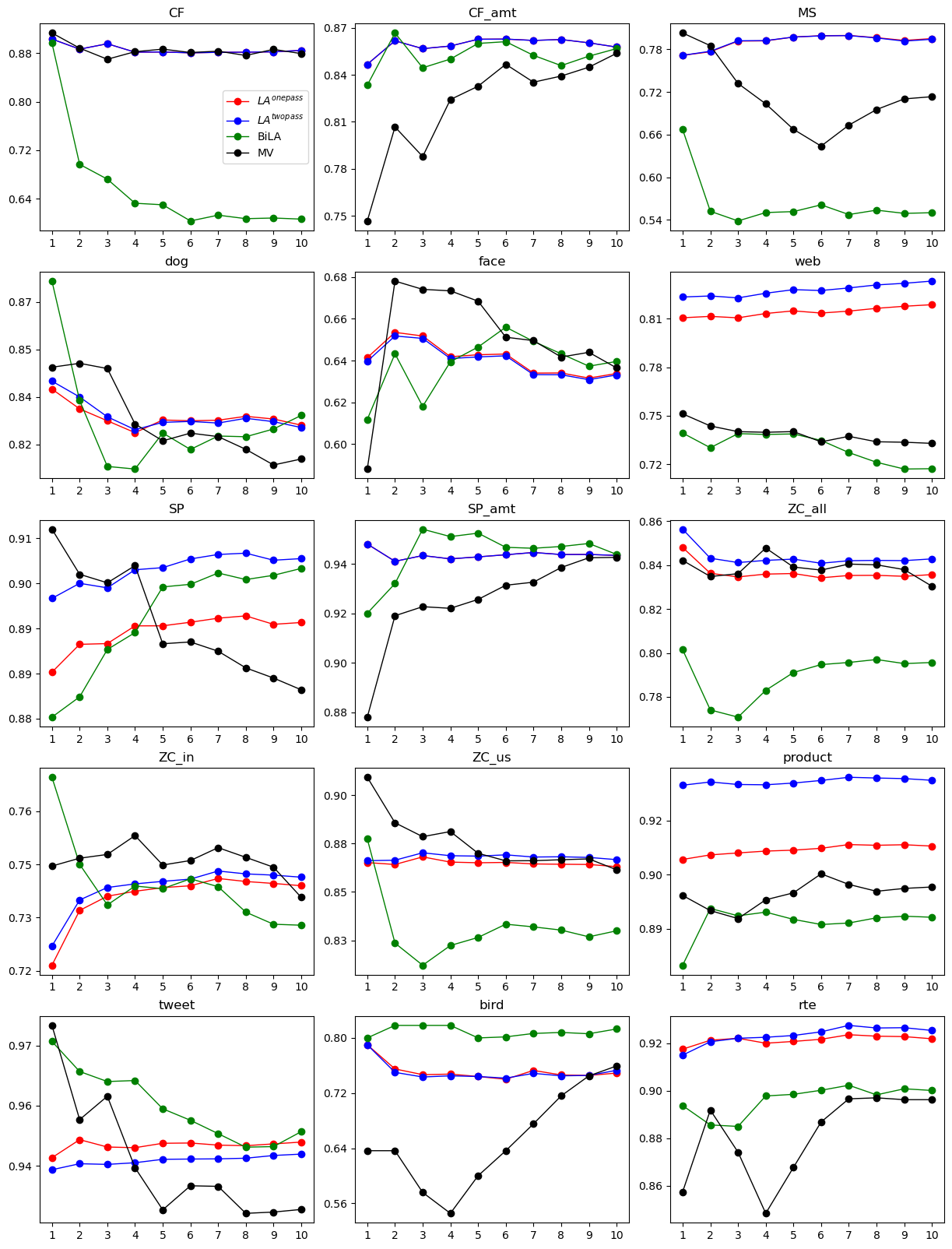}
	\caption{Online accuracies of each method performed on the selected datasets. The x-axis is the chunk index; the y-axis is the accuracy. The red and blue lines, representing the accuracies of LA\textsuperscript{twopass} and LA\textsuperscript{onepass}, may overlap in some datasets.}
\end{figure*}

\begin{figure*}[!ht]
	\centering
	\includegraphics[width=\textwidth]{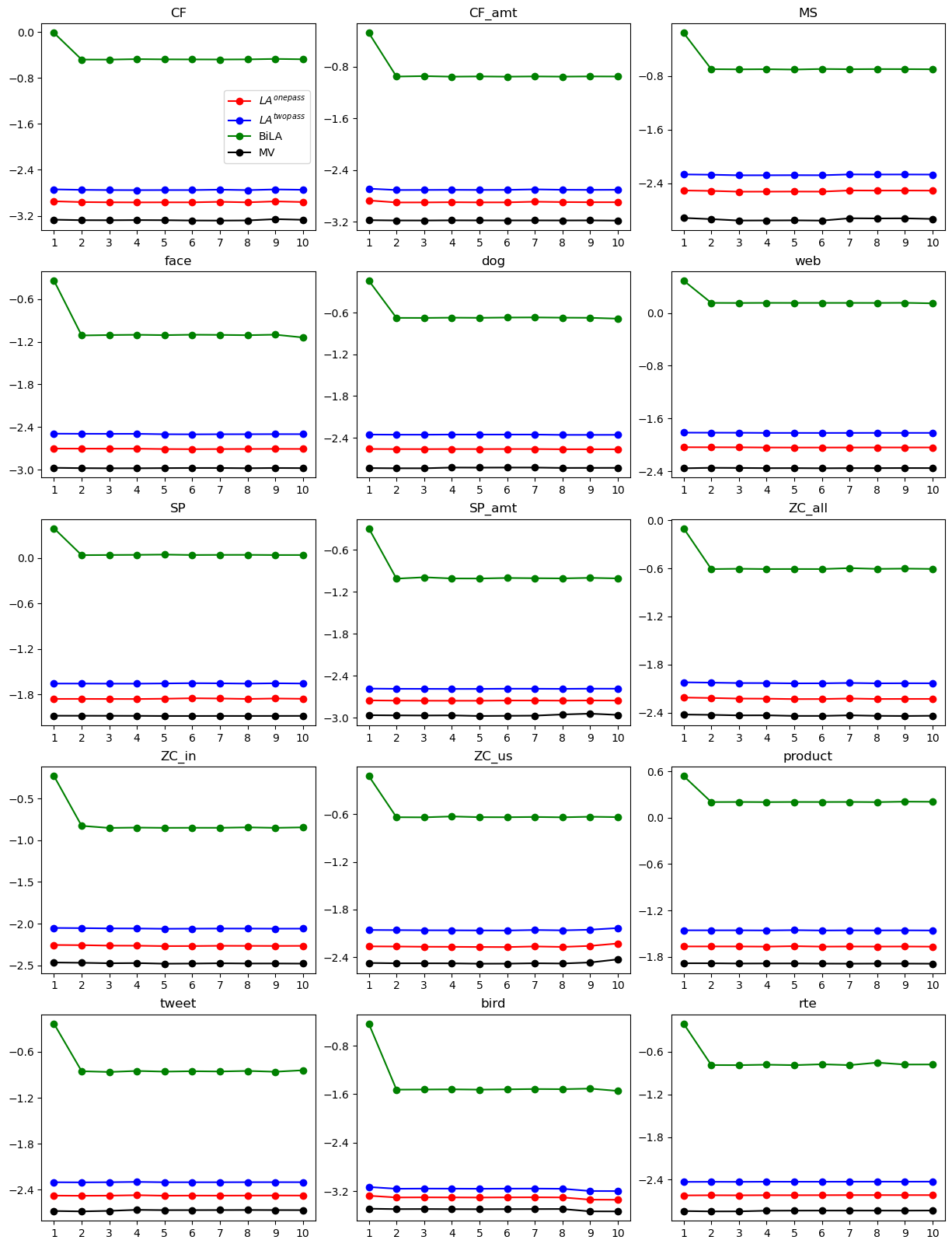}
	\caption{Online runtime ($\lg(sec)$) of each method performed on the selected datasets. The x-axis is the chunk index; the y-axis is $\lg(sec)$.}
\end{figure*}

\end{document}